\documentclass[conference]{IEEEtran}

\usepackage[utf8]{inputenc}
\usepackage[T1]{fontenc}
\usepackage{microtype}
\usepackage{cleveref}

\usepackage{todonotes}
\usepackage{tabularx}

\usepackage[keeplastbox]{flushend}

\usepackage{paralist}
\usepackage{booktabs}

\usepackage{graphicx}

\begin{document}

\title{Microservice Architectures for Advanced Driver Assistance Systems: A Case-Study}

\author{
\IEEEauthorblockN{Jannik Lotz and Andreas Vogelsang}
\IEEEauthorblockA{
Technische Universit\"at Berlin\\
\{jannik.lotz,andreas.vogelsang\}@tu-berlin.de}
\and
\IEEEauthorblockN{Ola Benderius}
\IEEEauthorblockA{
Chalmers University of Technology\\
ola.benderius@chalmers.se}
\and
\IEEEauthorblockN{Christian Berger}
\IEEEauthorblockA{
University of Gothenburg\\
christian.berger@gu.se}
}

\maketitle

\begin{abstract}
The technological advancements of recent years have steadily increased the complexity of 
vehicle-internal software systems, and the ongoing development towards autonomous 
driving will further aggravate this situation. This is leading to a level of complexity that is pushing the limits of existing vehicle software architectures and system designs.
By changing the software structure to a service-based architecture, companies in other domains successfully 
managed the rising complexity and created a more agile and future-oriented development process. 
This paper presents a case-study investigating the feasibility and possible
effects of changing the software architecture for a complex driver assistance function
to a microservice architecture. The complete procedure is described, starting with the
description of the software-environment and the corresponding requirements, followed by the
implementation, and the final testing. In addition, this paper provides a high-level evaluation
of the microservice architecture for the automotive use-case. The results
show that microservice architectures can reduce complexity and time-consuming
process steps and make the automotive software systems prepared for upcoming challenges
as long as the principles of microservice architectures are carefully followed.
\end{abstract}


\section{Introduction}
Service-oriented architectures and especially microservice architectures~(MSA) have been successfully applied to create flexible, maintainable, and scalable web applications and information systems.
As automotive software systems are becoming equally complex and critical to the development of modern cars, the MSA concept may have a positive impact on the architectures of such systems. Some potential advantages may be: (1)~Reuse of functionality, (2)~focus on data rather than actions, (3)~encapsulated and independent service behavior, (4)~continuous service integration and delivery, (5)~hierarchical in-vehicle function and software architecture, (6)~clear and explicit service documentation, and (7)~flexible service distribution to electronic control units~(ECUs) or cloud servers.

To have a closer look at the possible effects for the automotive industry by changing the software architecture style, we conducted a case study in the context of an advanced driver assistance system (ADAS) project. The developed system is close to an assistance feature in production and is designed, implemented, and tested as a microservice system. Shifting the architecture to the new style carries also some risk and challenges, ranging from whether an MSA is suitable for a system or project to avoiding overhead by distributing services in the wrong way. Potential solutions how to avoid these hazards are discussed in this paper. Lastly, the success of the MSA is coupled with technical progress, business aspects, and openness for the changeover, which can be created by a full understanding of the concept. Therefore, we provide a high-level architecture analysis of the potential strengths and opportunities but also the weaknesses and threats that come with this architectural style in the automotive use-case.

The main goal of this paper is to design, transform, implement, and test an ADAS close to series production into the state of the art microservice technique, which is used at Chalmers University of Technology. The case study project is based on an already existing lane detection algorithm, which is transformed and updated to the new OpenDLV microservice environment. The entire project implementation took place at the Division of Vehicle Engineering and Autonomous Systems~(VEAS) at the Chalmers University of Technology and the vehicle research laboratory Revere. The method how the ADAS was designed, implemented, and tested during the case study is described in this paper. Therefore, the existing services and components were analyzed, requirements were set up, and the system was implemented. Once the implementation phase was complete, the system was tested and the architecture was evaluated. In addition, background information about the concept, software environment, and standards are given.

\section{Background and Related Work}

Due to the high demands on reliability, functional safety, robustness, and resource efficiency for automotive systems, when for example compared to web applications, only a few architectural styles and patterns are used. Typically, most automotive software architectures can be considered \textit{component based}. In many cases, these components are interconnected that tightly that the architectures should be considered \textit{monolithic}.

There is work demonstrating how a monolithic application
can be transformed into a microservice system. For example, the experience
report from the banking sector dealt with the transformation of a currency conversion
system from Danske~Bank into a system based on microservices~\cite{Bucchiarone18}. Due to the enormous
size of the system, tasks such as fault tolerance mechanisms, concurrency handling, and
monitoring gained importance. Also the design of the system and the capability to
manage all services was a challenging task during the case-study. 

The software environment and workflow that is used during our case-study is based
on the developments and results from Berger~\textit{et al.}~\cite{Berger17}. 
Benderius~\textit{et al.}~\cite{Benderius18} have shown that OpenDLV 
in combination with Docker are a well suited software environment for a successful 
microservice deployment for vehicles and autonomous driving.

An alternative development environment is described
by Kugele~\textit{et al.}~\cite{Kugele18}. In their work, the throughput of the
OpenDDS software middleware is evaluated. Furthermore, a formal mapping of services
of the data distribution service (DDS) and a case-study about fail-operational behavior
are described. The paper concludes that the used SOA technique
in combination with DCPS is suitable for the automotive industry, but some points
regarding safety, certification, and security could not be clarified. Kugele~\textit{et al.}~point out that a commercial version of DDS could solve these problems. Also,
DDS in combination with the Docker platform is a suitable development method in an
agile work environment.

Another publish-subscribe middleware called Chromosome and a centralized
platform architecture for automotive applications, called Race, have similarities
with the above approach and with this work~\cite{Sommer13}. It is a central computing
architecture, which is supporting \textit{plug and play} on the software and network level. 

Kugele~\textit{et al.}~\cite{Kugele17} presented research challenges for a future-proof 
E/E~architecture, which were taken into account in the case-study and theoretical considerations
in this paper. In their paper, they discuss the relationship between future
E/E~challenges and the question of how SOA could help to overcome them based on
an interview study with twelve participants at BMW. In addition, Kugele~\textit{et al.} provide a formal
definition for the notion of a service in automotive software architectures.

\section{Study Design}
With our study, we aim to evaluate advantages and disadvantages of using containerized microservices to implement advanced driver assistance functions. We decided to perform this evaluation as a case study, in which we transform a lane following function as part of an existing vehicle software infrastructure and afterwards report our experiences.
To do so, we performed a SWOT-analysis based on our implementation and answer the following research question:

\textbf{RQ1: What are the strengths, weaknesses, opportunities, and threats of a microservice-based ADAS architecture?} 

Secondly, we were interested in how far best practices and anti-patterns from using microservices in other domains can be transferred to the automotive domain. Thus, we analyzed a study by Taibi and Lenarduzzi who provide a list of anti-patterns on projects that are based on the MSA principle~\cite{Taibi18}. Their paper is based on surveys and interviews with 72 experienced microservice developers from different industries about architectural smells, bad-practice, and anti-patterns.
We reflect this list of bad smells based on our study and analyze which bad smells are also relevant for automotive systems.

\textbf{RQ2: What are architectural smells that need to be considered in a microservice architecture for ADAS?} 

In the following section, we first report on the developed case, the environment, in which it has been realized, and how we implemented the microservices.

\section{Study Execution: An MSA-based Lane Following Function}

\subsection{Study Object}
The idea behind this paper is to transform an ADAS towards a microservice architecture. The
system should be close to a system that could be used in production vehicles. With
this intention, possible advantages, disadvantages, and characteristic of the MSA should
be determined. A lane following system is chosen, the modification and redesign is
promising to provide the desired view of the MSA technology. The core task of the system
is to detect and follow lanes on a normal road with street boundaries, where two lines are
marking the current lane. Based on the detected lanes, the system is able to calculate
a trajectory consisting of points, which are located in the center of the lane. This is
also the preferred trajectory the car will follow. To realize this, steering and acceleration requests
are sent with the help of a proxy representative service via messages on the regular
controller area network (CAN) vehicle bus. The resulting signals are then processed
by all addressed ECUs. There is no difference whether those signals are created by the
driver or a system.

\subsection{Study Environment and Constraints}
The case study implementation took place at the Division of Vehicle Engineering and Autonomous Systems (VEAS) at Chalmers University of Technology and the vehicle research laboratory Revere.
The implementation was subject to some constraints that were predefined by the study environment, as described below.

\subsubsection{OpenDLV and the libcluon Middleware}
OpenDLV is a modern microservice-based software ecosystem for self-driving vehicles. The microservice based software framework is open source and is designed to support the development, testing, and deployment of functions that are needed for autonomous systems such as ground vehicles and vessels. OpenDLV handles the hardware communication,
sensor fusion, and provides safety as well as other functionality essential to autonomous systems. 
The OpenDLV standard message set~(SMS) standardizes the communication in the
OpenDLV framework by decoupling the high-level application logic from the low-level
device drivers through well defined interfaces.
OpenDLV is built using the libcluon library\footnote{http://github.com/chrberger/libcluon}, which is the first and only single-file,
header-only middleware for distributed systems.

\subsubsection{Docker}
The open-source platform for container-based virtualization provides a series of tools for packaging, orchestration, and shipping of images.
Docker provides an isolated run-time environment, where users do not need deeper technical
knowledge about underlying levels such as device nodes, dependendent 3rd-party libraries, or further relevant software artifacts required to successfully execute an application. Furthermore, a public repository and cloud service, 
called Docker~Hub, allows every user to access, share, and download usable images.
The service also allows users to register externally triggered user-defined and containerized jobs
that can be used in an automated way to build, test, and deploy software.

Complex microservice systems are mostly composed of multiple encapsulated services
in distributed containers. With the help of the \textit{docker-compose} tool such 
multi-container applications can be defined, configured, run, and monitored throughout their entire life-cycle. All required
images are collected and specified in a YAML-formatted docker-compose file. Finally, a single 
command is sufficient to start or stop the complete system.

\subsubsection{Alpine Linux and the Linux Kernel Patch PREEMPT\_RT}
Alpine Linux is a Linux distribution that has the main focus on simplicity, security, and
resource efficiency. Hence, it is well fitted for images that are operating in embedded
systems. With the help of Alpine Linux a fully functional Docker image with all
necessary tools and libraries included typically result in a size of only 3--8~MB.  The
Preempt RT kernel patch allows time critical processes to interrupt lower prioritized processes at any time and to
occupy all resources to ensure that the timing constraints will be met. In addition,
the patch allows to lock specific resources for sensitive processes. It ensures that no
lower prioritized process can utilize the needed resources simultaneously. Real-time
patched Linux kernels are well tested and can be found in many embedded systems in
different application areas.

\subsection{Study Implementation}

\begin{figure}
\centering
  \includegraphics[width=\columnwidth]{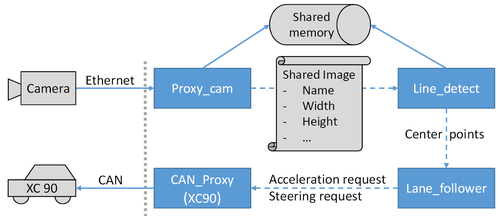}
  \caption{Architecture of the lane following system}
  \label{fig:systemOverview}
\end{figure}

The system implements a lane following function, which is already used at
the vehicle laboratory Revere, realized with the use of the open source library for computer vision and machine
learning, OpenCV. \figurename~\ref{fig:systemOverview} illustrates the design of the lane following driver
assistance function. The dashed arrows represent communication via the APIs, while
the solid arrows represent physical connections. The component \textit{shared image} shows
exemplary parameters that are needed to exchange data via a shared-memory. 
More details about the system and the used microservices, which are shown in the figure
as blue boxes, are presented hereafter. 

\textbf{The camera proxy microservice: }
The camera proxy microservice connects with any selected OpenCV accessible
camera device. It is responsible for the data input of the system and delivers a video
stream enriched with necessary metadata such as timestamps and sender id. In
a production-ready system, the camera would likely be accessed via an internal vehicle
bus. In this experimental set-up however, an Ethernet network camera is used, which is
accessible via the on-board Ethernet network built into the experimental platform Volvo XC90. According
to the microservice principles, any camera can be used as long as the input format, resolution,
and update rates are supported. The camera name, the name of the shared memory
section, the resolution, and other parameters have to be given as command line arguments
when starting the microservice.

The camera service updates the shared memory when a new frame is captured, and any consumer
service may momentarily lock and copy the data to local memory for further processing.
If the frame rate of the producer is faster than any consumer, the consumer will simply
lock and copy data as soon as ready (i.e.~processing will run at a slower rate, but will not affect
the rate of the producer), and if the consumer is faster it will simply wait for
data as the shared memory lock is blocking.

\textbf{The lane detection microservice: }
The fundamental lane detection algorithm was realized as a student project at the
Chalmers University of Technology. Later on, the algorithm was tuned and used
at the Chalmers Revere laboratory in different projects. In our case study, we further
transformed and adapted the algorithm to run as a microservice with OpenDLV as
the software environment. The algorithm uses different image processing methods from
the OpenCV library such as the Canny edge detection, intensity thresholds, and the Hough
transformations in order to detect lanes.

For the case study, we had to apply many changes to the image processing function.
Several parts had to be changed due to recent changes in OpenDLV and libcluon, including
datatype conversions, OpenDLV SMS compliance, timestamping, and initialization operations. 
During testing and tuning, relevant parameters are given as command line arguments
through the Docker-compose file. The return parameter of the image processing algorithm
are center points for the lane which should be followed and are located between
the detected road markings or road edges. This data is transmitted through UDP multicast,
which is the default transmission method of libcluon. The message specification is part
of the OpenDLV SMS.

\textbf{The lane follower microservice: }
The lane follower microservice calculates the steering angle and the acceleration request
given the current perceptual state of the vehicle, determined by using the center points of the
detected lane from the lane detection microservice. The lane follower is also checking
the validity of the detected lanes by comparing timestamps of messages at different
sections. This safety query could inform the driver and switch off the system as soon
as the images are no longer available on time. The actuation request messages sent from
the lane follower microservice is part of the OpenDLV SMS, and is standardized between
vehicles and vehicle types allowing for microservice reusability between platforms.

\textbf{The CAN proxy microservice: }
The CAN proxy microservice is the connection between the high-level OpenDLV software
components and the low-level software layers of the hardware platform. The requests for
steering and acceleration are converted to CAN-messages and sent on the associated bus
combined with other essential information and parameters. All necessary ECUs, which
are part of the powertrain and steering system, are addressed by message identifiers.
The ECUs process the incoming data and calculate new states. The computed state
consist of different request which are executed by the corresponding vehicle actuators.
These requests and messages do not differ from the regular operating state, in which
they are controlled by a human driver. Therefore, the signals can be processed the same way down
the line.

\section{Study Results}
\subsection{RQ1: Strengths, Weaknesses, Opportunities, and Threats}
We evaluate our implementation based on a SWOT analysis. 
Table~\ref{tbl:SWOT} provides an overview of the aspects that we detail in the following:

\begin{table}
\renewcommand{\arraystretch}{1.3}
\centering
\caption{Summary of the evaluation presented as a SWOT-matrix}
\label{tbl:SWOT}
\begin{tabularx}{\columnwidth}{X|X}
\hline
\textbf{Strengths}&\textbf{Weaknesses}\\
Modularity and reuse & Shifting complexity\\
Scalability & Change of team structure\\
Independent development, deployment, testing & Introduction costs\\
\hline
\textbf{Opportunities}&\textbf{Threats}\\ 
Shorter time to market & Smells and anti-pattern\\
Mastering rising complexity & Security\\
Reduced dependency errors & Technology backlog\\
Redundant system deployment & \\
\hline
\end{tabularx}
\end{table}

\subsubsection{Strengths}

\textbf{Modularity and Reuse}
One of the core characteristics of microservices leads directly to one of the most valuable
strengths. Since microservices are containerized and communicate only via messages or through
specified APIs, the underlying hardware is secondary. In addition, each service can be
combined as needed to create new systems or to support existing ones. 
Another benefit of modularity is the fact that single points of failure are easier
to avoid. When the architect implements services with resilience in mind, the system is
still functional if a service breaks down. Since most automotive functions are realized in
layered or component-based architecture style, this benefit is not completely new, but
is further strengthened and can be simplified by the MSA. 
The characteristics of high modularity could also benefit the trend of vehicle features
being activated and installed over the Internet. On a technical level this is realized with standard
ECUs and components, which are built into every production vehicle from scratch.
Through downloading and installing the software and adding this component to the
overall vehicle system, the feature will be activated. The electric car manufacturer
Tesla is already using this process to add new or delayed features to cars already sold.
Because of the small size, clearly defined APIs, and the resulting modularity, microservices
are a good match to cover the software part of this process.

\textbf{Scalability}
Cloud computing, big data processing, and artificial intelligence are becoming increasingly
important within vehicle systems. Due to the amount of data or poor network
quality however, not all of this operation can be processed externally. For this reason, it
would be desirable to have unused computing resources on demand. To achieve this, it
would be necessary for the software to be able to adapt to the needs of both the vehicle
and the driver, especially in the event of bottlenecks. For this purpose, containerized
microservices could be cloned automatically and distributed to
ECUs, which are not fully utilized. For this task, a load balancing unit could be integrated
into the system with little effort. Containerization makes it possible to clone a Docker
image and call the additional service from the Docker-compose file without further modifications. The scalability was also proven during the vehicle test. For this purpose,
the lane detection service was cloned and both containers were executed simultaneously.
The same method can also be applied to the external back-end, where individual
services can be scaled up or down depending on the system load. The strength of scalability leads to a higher degree of utilization of the available resources, which results in cost savings.

\textbf{Independent development, deployment, and testing}
In an MSA, each service can be developed and deployed by independent teams. 
Since the deployment units are smaller and independent from each other, every service
can be built, tested, and deployed without affecting the system in place. In addition, containerized
microservices offer the possibility to compare different versions of a service,
by simply using the same input data and by monitoring the output. 
Another benefit of small software modules is that parts of the deployment workflow could be automated, versioned, and again developed independently of other services, which leads to a faster integration of software functions or updates. This strength was also validated during the case study. Existing services could easily be accessed via the Revere source code repository, while other services could be changed and developed independently. In addition, it was
possible to test different versions of the individual services without having to apply
major changes to the overall system structure.

\subsubsection{Weaknesses}
\textbf{Shifting complexity:}
One problem often found by changing the software architecture to microservices
according to different experience reports is that complexity shifts
from module-level to a higher level~\cite{Bucchiarone18}. The challenge of creating, controlling, and monitoring
the system is increasing. The larger the number of individual services active in a system,
the more complicated it is to keep track. In addition, the test complexity shifts, since
services are smaller and contain only a sub-task which leads to simpler module tests.
In addition, comparing the output with the same input data makes it easier to test
different versions of a module. On the other hand, however, it is much more challenging
to test the interaction between dozens of services at the integration or system test level.
In summary, the complexity is shifting from the individual module complexity to
a higher level, where architectural design, interaction between services, and monitoring takes place~\cite{Savchenko15}.

\textbf{Change of team structure:}
Old business structures must be changed to enable successful service development. This is typically a major challenge for most established car manufactures. Structures and responsibilities are usually divided according to
the technical vehicle domains up to board level. This division was created in the past,
when a clear separation was still possible. Furthermore, project management, supplier contracts, and budget
planning are based on these structures. If restructuring is only partial or is still
directed to the domains, the MSA will not be able to take advantage of the benefits. It
needs freedom of action, willpower as well as ambition in order to break these structures
and make full use of the MSA advantages and to simplify inter-domain communication
and accelerate development. Furthermore, a team responsible for a service must cover the
entire life cycle of the product. It follows that each team needs knowledge
about the entire product cycle. This structure is not common in traditional automotive
software-teams and it requires courage to transform teams adapted to the microservice
principles.

\textbf{Introduction costs:}
Fundamental technology changes are usually associated with high costs. For example,
microservices and containerization come along with an overhead in terms of memory consumption.
Since the ECUs used for ADAS usually have to handle complex tasks, they
are among the most powerful in a vehicle. However, when planning a new microservice
system, it must be taken into account that MSA usually requires more memory and
computing power for communication processing. In addition to the possible hardware
modifications, cost of structural changes will affect manufacturing. It takes resources
to restructure existing teams, contracts, and business structures. Furthermore, since the
MSA advise that teams cover the entire product cycle, the resulting cost can no longer
be directly allocated to the conventional cost centers. This could lead to an MSA system
style to be more expensive in the beginning. The majority of these costs are attributable
to the introduction, personnel training, and restructuring.

\subsubsection{Opportunities}
\textbf{Shorter time to market}
As mentioned above in the introduction, it is the time to market to be shortened to
keep track with innovations and competitors. Customers are no longer satisfied with
only experiencing a new vehicle functions or a new software system when purchasing a
new vehicle generation. Since microservices are modular, they provide optimal
conditions for simplified and accelerated update procedure. The prerequisite for this is that the vehicle is
equipped with over-the-air update functionality and connection to the Internet.

\textbf{Reduced dependency errors}
In most vehicle software architectures, such as component-based architectures, there
is a multitude of dependencies between functions and sub-functions. This is however
not only in contrast to the intention to design internal vehicle systems as modular as
possible, but is also very error-prone, since changes to a sub-function can have effects
on other elements. Vogelsang and Fuhrmann discussed this problem~\cite{Vogelsang13}, 
showing that the high degree of dependency leads to that a function
developer is only aware of about 50\% of the dependencies. 
While this is also a problem for microservice architectures, there are already available solutions developed and applied even at the scale of, for example, Netflix and Spotify.

\textbf{Master rising complexity of inter-domain systems}
The case study has shown the procedure to split the function complexity
among services. The result is an ADAS that consists of four domain-independent
microservices. Each service has a fixed task that corresponds to a capability. In a monolithic
approach, the entire function would have been handled in one large component.
Even if the system had been implemented with a component-based architecture, there
would be strong dependencies between the different sub-components. Thus, the complexity of
the inter-domain basic function was significantly reduced with the MSA.

\textbf{Redundant system deployment}
Autonomous driving includes a high number of safety critical systems. Even when a
subsystem of a function like steering or braking fails, the vehicle has to be able to
manage the situation without risking the life of the passenger or other road users. 
Microservices provide an opportunity to clone the relevant services to an identical hardware to detect deviations and errors. A logic component can then decide, which alternative is more trustworthy and
can ensure continued system operation.

\subsubsection{Threats}
\textbf{Microservice smells and anti-patterns}
When MSAs are used, they should be checked for sub-optimal design and potential sources of undesired behavior, so called architectural smells.
These smells may lead to increased need of resources and in worst case could lead to the failure of the entire project.
This could pose a threat to the new architecture, especially in pilot projects, which are
often used to demonstrate the potential and feasibility of new technologies.

\textbf{Security}
Since the system is more loosely coupled and composed of many services with many
open interfaces, the surface for attacks is increasing. In traditional architectures,
architects and developers had to secure systems, which consisted of only a few elements.
It is different with systems that are created with the component-based architecture,
where it also occurs that a system consists of many components. Since components are
units of functionality and not containerized, they are, however, generally more tightly
coupled~\cite{Mustafa18}. In both cases, it is comparatively easier to keep track of the entries for a
possible attack. On the other hand, however, when a component within a traditionally
designed system breaks down through an attack, it is very likely that the whole functionality
collapses. This scenario can be mitigated by the MSA by using the architecture,
every developer has to be more conscious of security and being aware of attack scenarios.

\textbf{Technological backlog}
To unlock its advantages, the MSA depends on further developments and introducing
of supporting technologies. The main drivers thereby are Ethernet for the internal
data transfer, the fifth generation~(5G) of cellular mobile communications for external
data exchange, and more powerful ECUs.
Vehicles need to exchange an ever-growing amount of data to update the internal software,
release bug-fixes, process data, and provide vehicle-to-everything~(V2X) connectivity.
Hence, communication channels with high data throughput and area-wide network
expansions are necessary. In addition, high-speed data exchange is required
as some information must be transmitted in real-time. 
Driven by the entry of computing power-consuming image processing, pattern recognition,
and other AI functionality, ECU hardware becomes more powerful nowadays.
This trend will extend to other domains in the future. Since the memory or computing overhead
of the architecture is low compared to functions such as image processing or
AI functionality, software architectures with overhead benefit from this progress and it
becomes easier for the architect to use a service-based software architecture. However,
without the simultaneous development and broad introduction of these technologies
into series production, MSA will not be able to exploit all advantages. The effort
for changing the architecture style would probably not be worth it.

\subsection{RQ2: Architectural Smells}
Taibi and Lenarduzzi provide a list of anti-patterns in projects that are based on the MSA principle~\cite{Taibi18}. Their paper is based on surveys and interviews with 72~experienced microservice developers from different industries about architectural smells, bad practice, and anti-patterns.
The identified microservice smells are not completely applicable to automotive software
or embedded systems. 
Table~\ref{tbl:smells} provides an overview of the smells and whether they also applied to the case discussed here.

\begin{table}
\centering
\caption{Smells}
\label{tbl:smells}
\begin{tabular}{lp{1.7cm}}
\toprule
\textbf{Smell} & \textbf{Relevant for \newline automotive} \\
\midrule
Hard-coded endpoints & no\\
ESB usage & no \\
Not having an API gateway & no\\
Too many standards & no\\
Cyclic dependency & yes\\
Wrong cuts & yes\\
Shared persistency & yes\\
API versioning & yes\\
Inappropriate service intimacy & yes\\
Shared libraries & yes\\
Microservice greedy & yes \\
\bottomrule
\end{tabular}
\end{table}

\textbf{Hard-Coded Endpoints:}
This smell occurs if hard-coded IP addresses are used to connect microservices.
This smell did not apply to our case study because the technical infrastructure in vehicles differs from web or information systems.
In both setups, messages are transferred via signals over communication
systems. The difference is in the method of message transfer. Most automotive
systems use broadcasting as the method for transferring messages. This means, to
execute the required command, every message is provided with an identifier to address
the corresponding ECU and is sent to all ECUs. Then, each microcontroller
decides depending on the identifier, whether the message is processed or dropped.
In web or information systems, direct message with hard-coded IP addresses and ports can be used. 
Since mostly broadcast messages are used in automotive systems, hard-coded endpoints rarely occur.

\textbf{ESB Usage:}
This smell occurs if microservices communicate via an enterprise service bus (ESB).
An ESB is a communication method between mutually interacting software
applications. The ESB is providing a central place where services, application,
and resources can be connected.
Due to the automotive specific ECU distribution and communication mechanism
(broadcasting), it would require an enormous amount of effort to integrate
such a central controller. Besides that, it would be a single point of failure component.
For this reason, this smell does not need to be taken into account.

\textbf{Not having an API Gateway:}
The data needed by consumers is usually different from the data
provided by the various APIs of each service. In a worst case scenario, the service
consumers would communicate directly with different microservices, increasing
the complexity of the system and decreasing its ease of maintenance.
An API gateway is the entry point for all clients and is handling all
requests. This makes it easier to orchestrate, monitor, and secure the system. Automotive
companies will face with this problem less as most hardware
components and interfaces associated therewith have already been determined
at the planning stage of the vehicle. In addition, manufacturers have more
decision-making power over the devices they use within the car, which leads to the
fact that the variety of access types and devices is smaller than in the web domain.
Furthermore, an API gateway is like the ESB a potential single point of failure
component and will increase the system overhead concerning computing-power.

\textbf{Too Many Standards:}
This smell occurs if different development languages, protocols, frameworks, and the like are
used.
Due to strict definitions, standards, and requirements during the development of
a vehicle, the scenario is very limited. Furthermore, microservice frameworks
such as OpenDLV often provide basic conditions and rules for the planning and
design of a system. Especially for safety critical systems, this smell is not very relevant.

\textbf{Wrong Cuts:}
This smell occurs if microservices are split based on technical layers (e.g.~presentation,
business, and data layers) instead of business capabilities.
As a result, wrong separation of concerns and increased data-splitting complexity
occurs.
The prospected solution to that is to perform a clear analysis of business processes and the need for resources.

The capabilities in the car industry are mostly separated along domains, for example:
Powertrain, chassis, comfort, entertainment systems, etc. Embedded systems that run in just one of those domains should barely suffer from this problem because responsibilities are clearly separated. In addition, the teams are mostly independent and, except for data exchange, no cross-functionality is required. This clear separation is
becoming increasingly blurred, so a new challenge is that functions are increasingly
being distributed across domain boundaries. For this type of functions, it is important to split the responsibilities within the departments, according to the capability, to reduce potential complexity and error sources. Cross-domain communication and team structures are needed to ensure that a team can work autonomously
and is able to focus on the value of the system. Traditional team borders and responsibilities have to be softened
and reassembled to make the microservice approach successful in the vehicle industry.
For projects like our case study, competences from different research domains
need to be grouped together. The first services can be assigned to the perception
area followed by the logical part, in which the image-processing takes place.
In the end, action services receive the commands and transfer them into state changes
of the vehicle. This processing chain can only be successfully implemented and deployed
when the work is independent from other projects and with clear responsibilities
according to capabilities.

\textbf{Cyclic Dependency:}
This smell occurs if a cyclic chain of calls between microservices exists.
Microservices involved in a cyclic dependency are difficult to spot at first place and hard to maintain or to
reuse in isolation.
A solution is to refine the cycles according to their shape and apply the API gateway
pattern.

Due to the reasons mentioned above at the point \textit{not having an API gateway}, to use
an API gateway is not an optimal option to prevent this smell in vehicle-systems. This
relation between modules exists also in other software architectures and can usually be
avoided by a smart and well thought through system architecture and design. For instance,
by trying to analyze the data flow, possible cycles could be detected and refined. This
smell is well known at Revere and is prevented by trying to design the system in such
a way that a directed flow is always clearly identifiable. Moreover, a cyclic loop is
prevented at Revere by the requirement that a microservice should never receive its
own message.

\textbf{Shared Persistency:}
This smell occurs if different microservices access the same relational database. In the worst
case, different services access the same entities of the same relational database.
This smell results in strong coupling between microservices connected to the same data,
reducing team and service independence.
There are three possible solutions to this smell: (1)~use independent databases for each service,
(2)~use a shared database with a set of private tables for each service that can be accessed
by only that service, or (3)~use a private database schema for each service.

Mainly due to cost pressure in the automotive domain, it is not always possible to give each microservice
its own storage data-space. It must be carefully considered in which case a common
data-space could be used. On the other hand, a separated and private data area
could give advantages to highly safety-critical systems. A service may be cloned
multiple times or just scaled and split on several ECUs with own databases. In addition, the processing power of ECUs, especially in inter-domain functions, is constantly rising. All these individual requirements lead
to the fact that the proposed solutions cannot be assigned completely and have to be
adapted partly for the automotive sector. To summarize, the system components, which
could lead to the shared persistency anti-pattern, have to be reviewed strictly during
the system design phase and after coupling of microservices. For the case-study at
Revere, the shared-memory technique with a lock mechanism is used to reduce a potential risk of this microservice smell. Since the Volvo XC90 is a test vehicle, it is well-resourced with additionally computing power. Thus, a possible resource constraint played a subordinate role in the case study.

\textbf{Shared Libraries:}
This smell occurs if different microservices share libraries.
In this case, microservices are tightly coupled, leading to a loss of independence
between them. Moreover, teams need to coordinate with each other when they need to
modify the shared library.
There are two possible solutions: (1)~accept the redundancy that increases the dependency among teams, or
(2)~extract the library to a new shared service that can be deployed and developed
independently by the connected microservices.

For this smell, both solutions are transferable and should be considered in
exceptional circumstances, for example, in data-space critical environments. However,
solution 1 should be considered even more carefully as the shared library could result
in a single point of failure. In the worst case, this means that a faulty
library can have negative effects on several services and subsystems. In addition, the
Docker environment advises that each container should be as independent and self-contained as possible.
By fully integrating the required libraries into each independent Docker image, we avoided shared
libraries during our case-study.

\textbf{API versioning:}
This smell occurs if APIs are not semantically versioned. 
In case of new versions of non-semantically versioned APIs, API consumers
may face connection issues. For example, the returning data might be different
or might need to be called differently.
Therefore, APIs need to be semantically versioned to allow services to know whether
they are communicating with the right version of the service or whether they need to
adapt to a new contract.

The Revere team is using Git as a tool for version control, and the Git commit hash
key is used to keep track of the code version. This key is also used in the Docker
image tag to achieve traceability between the current version of source code and deployable
software bundle which also includes the API-version. This naming procedure
also improves the overview for frequent testing and for comparing different versions. To
integrate the lane detection system into the existing Revere code base, the procedure
for version control was adopted for the case-study.

\textbf{Inappropriate Service Intimacy:}
This smell occurs if a microservice keeps on connecting to private data from other services
instead of dealing with its own data, which increases coupling between
microservices or, in the worst case, might introduce unwanted couplings or dependencies, that are not explicit.

This architectural smell is often a hint that two services are too tightly coupled.
One goal of a good MSA is to minimize dependencies between services by well designed interfaces and service contracts to be transferable to different systems and to cooperate with various versions. However, it is a balancing act between too few and too many services, which will be described in the next microservice smell.
In general, single functions in automotive systems do not store much data, as they are mainly reactive and data-driven. However, a common (bad) practice is to use global variables instead of properly defined interfaces to access other function's internal state or data. This is an instance of the Inappropriate Service Intimacy smell. Therefore, we did not use any global variables in our case study.

\textbf{Microservice Greedy:}
This smell occurs if teams tend to create new microservices for each feature, even when they
are not needed.
This smell can generate an explosion in the number of microservices, resulting in a huge system that is hard to maintain because of its size.

In the automotive industry, costs and resources per unit for engineering and development of a car play
an important role. Therefore, architects and software developers are anxious to save
resources and should try to reduce unnecessary elements, where ever it is possible. This
circumstance weakens the microservice greedy smell for the automotive use-case. The
design and structure of the system are well coordinated with experienced microservice
architects, which results in a manageable number of services. Therefore, there was no
risk of getting into this bad-practice during the implementation of our case study.

\section{Conclusions}
The goal of this paper was to examine whether the MSA can be used as a technically advanced
and problem solving software architecture in the automotive industry. We conducted a case study to promote understanding and examine the feasibility of the architectural change. A modern development and test environment was available at Chalmers University of Technology and the Revere research laboratory to conduct the case study.

The lane following ADAS system worked as planned with the microservice
architecture and the specified requirements were met. Like in other software
architecture styles, there are code or architectural smells in the MSA that could trigger
solution patterns that do not benefit the project development. We described smells,
the corresponding problem that may arise, and possible solutions as well as additional adaptations to the automotive use-case. Finally, we performed a high level evaluation of the MSA to identify the strengths, weaknesses as well as potentials and threats to the automotive industry. At several points of the analysis, experience and observations from the case study were included.
The outcome of the case-study shows that the architecture in combination with a
simple and powerful software environment is a promising approach to the automotive
industry for mastering current challenges and to be prepared for upcoming tasks. The
MSA is advantageous in areas, where maintainability of source code, easy scalability, and high modularity are required. However, project architects must be aware that they must individually evaluate the software architecture for each project in terms of usage, size, and scope. Service-based approaches are not a panacea for all software
related problems in the automotive industry. Furthermore, the architect must be aware
of possible weaknesses and threats in the changeover of the software architecture to
microservices. After all, both, architects and software developers, should keep
an eye on any smells within source code that could cause problems. Once these
points are not only considered, but actively refactored, the MSA is ready for successful
application in complex ADAS and autonomous driving functions.

\bibliographystyle{IEEEtran}
\bibliography{bib}
\end{document}